\documentclass[a4paper,amsmath,amssymb,prl,twocolumn,superscriptaddress]{revtex4-1}
\usepackage[T1]{fontenc}
\usepackage{graphicx}
\usepackage{dcolumn}
\usepackage{bm}
\usepackage{latexsym}
\usepackage{epic,eepic}
\usepackage{subfigure}
\usepackage{overpic}
\usepackage[retainorgcmds]{IEEEtrantools}
\usepackage{hyperref}
\usepackage{grffile}
\usepackage{microtype}
\usepackage[super]{nth}

\def\kcore{$k$-core}

\def\ksubtree{$(k-1)$-ary subtree}
\def\kisubtree{$(k_i-1)$-ary subtree}
\def\hkcore{heterogeneous $k$-core}
\def\hkcore{HKC}

\def\S{\mathcal{S}}

\def\er{Erd\H{o}s-R\'enyi}
\def\k{\mathbf{k}}

\DeclareGraphicsExtensions{.png,.pdf}

\begin{document}
\title{Tricritical point in heterogeneous $k$-core percolation}
\author{Davide Cellai}
\affiliation{MACSI, Department of Mathematics and Statistics, University of Limerick, Ireland}
\author{Aonghus Lawlor}
\affiliation{CBNI, University College Dublin, Belfield, Dublin 4, Ireland}
\author{Kenneth A.~Dawson}
\affiliation{CBNI, University College Dublin, Belfield, Dublin 4, Ireland}
\author{James P.~Gleeson}
\affiliation{MACSI, Department of Mathematics and Statistics, University of Limerick, Ireland}

\begin{abstract}
$k$-core percolation is an extension of the concept of classical
percolation and is particularly relevant to understand the
resilience of complex networks under random damage. A new analytical
formalism has been recently proposed to deal with heterogeneous
$k$-cores, where each vertex is assigned a local threshold $k_{i}$. In
this paper we identify a binary mixture of heterogeneous $k$-cores
which exhibits a tricritical point. We investigate the new scaling
scenario and calculate the relevant critical exponents, by analytical
and computational methods, for Erd\H{o}s-R\'enyi~networks and 2d
square lattices.
\end{abstract}
\date{\today}
\maketitle

Tricritical points (TCPs) are interesting critical phenomena in
statistical physics, constituting a natural switch between \nth{1} and
\nth{2} order phase transitions.
In other words, a TCP affords the possibility to smoothly control the
order of the phase transition by tuning the appropriate
parameter. 
In
the language of critical phenomena, this is equivalent to identifying
the position of a phase transition from the extrapolation of the order
parameter- while usually impossible in \nth{1} order transitions, it
can be done in the case of continuous ones \cite{vespignani2010}. In
the case of percolation models, there has recently been an attempt to
identify a TCP in a model mixing elements of classical and explosive
percolation in a lattice \cite{araujo2011,*achlioptas2009}, although there is now
evidence that the explosive percolation transition is continuous
\cite{dacosta2010,*grassberger2011,*riordan2011}. Other recent models
which allow control of the order of the transition include explosive
percolation on scale free networks \cite{radicchi2009}, and dependency
groups on interdependent networks \cite{parshani2010}. In this paper
we establish, for the first time, the presence of a TCP in a simple
extension of classical percolation, namely heterogeneous $k$-core
({\hkcore}) percolation, which has the advantage of a sound analytical
approach on random and complex networks
\cite{dorogovtsev2006,baxter2011}. We show analytical evidence of a
TCP in {\er} graphs and numerically we find similar phase diagram
topology in the square lattice. Finally, our model
appears to be in the same universality class of a model which
reproduces non-trivial signatures of liquid-glass transitions, including the
higher-order glass singularity
predicted by mode-coupling theory \cite{sellitto2010}.

A $k$-core is defined as the maximal network subset which survives
after a culling process which recursively removes all the vertices
(and adjacent edges) with less than $k$ neighbors.  As a
generalization of the concept of the giant component, the {\kcore}
gives a deeper insight into the structure and organization of complex
networks. It has been thoroughly investigated on Bethe lattices
\cite{chalupa1979}, random graphs
\cite{Balogh:2006p4406,dorogovtsev2006} and, using a numerical
approach, on various types of lattices
\cite{Adler:1989p5803,*branco1999}. The \kcore~percolation analysis
has found several applications in varied areas of science including
protein interaction networks \cite{wuchty2005}, jamming \cite{schwarz2006}, neural networks
\cite{Chatterjee2007}, granular gases \cite{alvarez2007} and evolution
\cite{klimek2009}.
Important insights into the resilience of networks under damage
\cite{cohen2000,*gleeson2008} and spreading of influence in social
networks \cite{kitsak2010} are gleaned from an understanding of the
\kcore~structure of the network.  As in
Ref.~\cite{dorogovtsev2006,goltsev2006}, we can study {\kcore}
percolation on networks after randomly removing a fraction $1-p$ of
vertices.
We use the treelike properties of the configuration model
\cite{newman2003}, in which the number of loops vanishes as
$N\to\infty$, which guarantees that if a {\kcore} exists, it must be
infinite, at least if $k\geqslant2$
\cite{chalupa1979,dorogovtsev2006}. In the {\hkcore} extension
\cite{baxter2011} each vertex has its own threshold and the culling
process is based on local, vertex-dependent rules. Although Baxter
\emph{et al.} developed results for an arbitrary distribution of
vertex thresholds, they study binary mixtures of vertices of types $a$
and $b$, with thresholds $k_{a}=1$, $k_{b}\geqslant 3$. The first
heterogeneous models of this kind were investigated by Branco
\cite{branco1993} on a Bethe lattice, whereas the related problem of
bootstrap percolation (BP) has been much studied on regular lattices
\cite{Adler:1989p5803,DeGregorio:2004p1889}.  Here we focus on the
case $\k\equiv(k_a,k_b)=(2,3)$.

We start with a binary mixture ($k_a$,$k_b$), where vertices have been
randomly assigned two thresholds $k_a$ and $k_b$ (say $k_a<k_b$) with
probability $r$ and $1-r$, respectively. Finite clusters are a
possibility when $k_a=1$ and so we must make a distinction between
$M_{ab}$, the probability that a randomly chosen vertex belongs to the
{\hkcore}, and $\S_{ab}$, the probability that a randomly chosen
vertex belongs to the \emph{giant} component of the {\hkcore}. We will
show that in the case $\k=(2,3)$ these two quantities are coincident,
but there are relevant examples where they are not \cite{baxter2011}.

In the original {\kcore} formalism, given the end of an edge, a
{\ksubtree} is defined as the tree where, as we traverse it, each
vertex has at least $k-1$ outgoing edges, apart from the one we came
in. Instead, considering a {\hkcore}, every vertex $i$ may have a
different threshold $k_i$. The {\kisubtree}, then, is the tree in
which, as we traverse it, each encountered vertex has at least $k_i-1$
child edges. We define $Z$ as the probability that a randomly chosen
vertex is the root of a {\kisubtree}. Taking advantage of the local
treelike nature of the configuration model, $Z$ is related to $M_{ab}$
as \cite{baxter2011}:
\begin{IEEEeqnarray}{rCl}
	M_{ab}(p)  &=&  \bar{M}_a(p) + \bar{M}_b(p) = p r\sum_{q=k_a}^{\infty} P(q) \Phi_{q}^{k_a}(Z,Z) +\nonumber\\
	&& + p(1-r)\sum_{q=k_b}^{\infty} P(q) \Phi_{q}^{k_b}(Z,Z)
	\label{eq:mass-k}
\end{IEEEeqnarray}
where $\bar{M}_{a(b)}(p)$ is the fraction of nodes of type $a(b)$ in
the \hkcore, respectively, $P(q)$ is the degree distribution and we
have used the convenient auxiliary function:
\begin{equation}
	\Phi_{q}^{k}(X,Z) = \sum_{l=k}^{q} {q \choose l} (1-Z)^{q-l} \sum_{m=1}^l {l \choose m} X^m (Z-X)^{l-m} \nonumber.
\end{equation}
The quantity $\Phi_{q}^{k_{a(b)}}(Z,Z)$ in (\ref{eq:mass-k})
represents the probability that a vertex of type $a(b)$ of degree $q$
has at least $k_a(k_b)$ edges which are roots of a \kisubtree. This
quantity is summed over all possible degrees, taking account of the
relevant vertex type fraction. The self-consistent equation for $Z$
is:
\begin{IEEEeqnarray}{rCl}
	Z &=& p r \sum_{q=k_a}^{\infty} \case{qP(q)}{\langle q \rangle} \Phi_{q-1}^{k_a-1}(Z,Z) +\nonumber\\
	&& + p(1-r)\sum_{q=k_b}^{\infty} \case{qP(q)}{\langle q \rangle} \Phi_{q-1}^{k_b-1}(Z,Z)
	\label{eq:Z-equation}
\end{IEEEeqnarray}
We now consider the probability $X$, that a randomly chosen edge leads
to a vertex which is the root of an \emph{infinite} \kisubtree. In the
case of a binary mixture, $X$ is written as \cite{baxter2011}
\begin{IEEEeqnarray}{rCl}
	X &=& pr \sum_{q=k_a}^{\infty} \case{qP(q)}{\langle q \rangle} \Phi_{q-1}^{k_a-1}(X,Z) +\nonumber\\
	&& +p(1-r)\sum_{q=k_b}^{\infty} \case{qP(q)}{\langle q \rangle} \Phi_{q-1}^{k_b-1}(X,Z),
	\label{eq:X-equation}
\end{IEEEeqnarray}
The fraction of vertices in the giant \hkcore~$\S_{ab}$, then, is given by 
$\S_{ab}(p) = \bar{\S}_{a}(p) + \bar{\S}_{b}(p)$,
where the fraction of nodes of type $a$ is $\bar{\S}_{a}(p) = p
r\sum_{q=k_a}^{\infty} P(q) \Phi_{q}^{k_a}(X,Z)$ and an analogous expression holds for $\bar{\S}_{b}(p) $.
\begin{figure}[bht!]
	\begin{overpic}[width=\columnwidth]
          {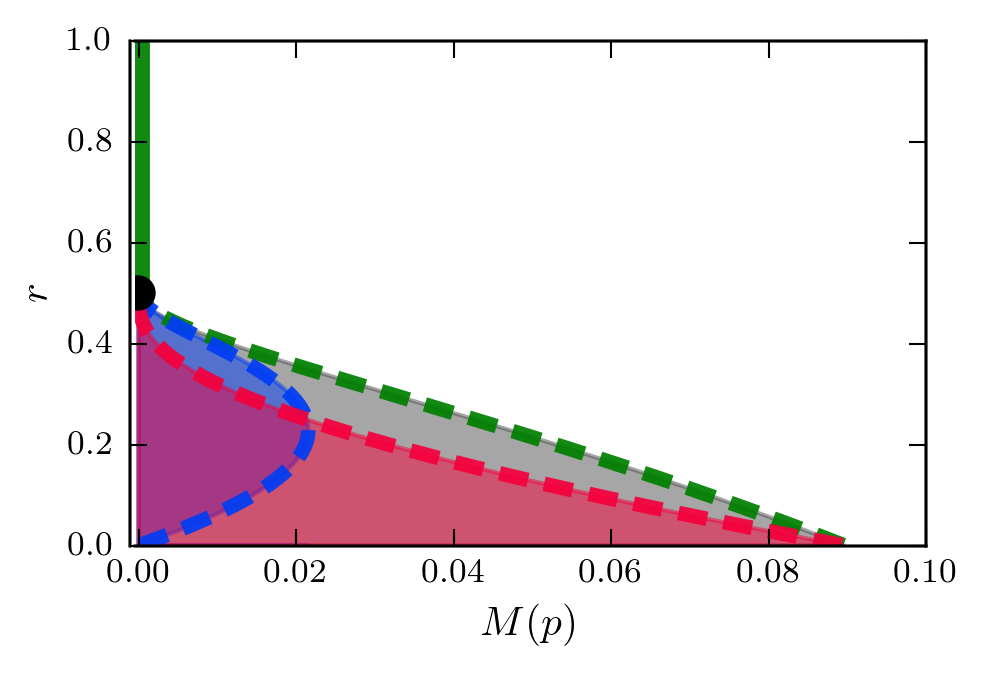}
		\put(61,24){\includegraphics[height=0.4\columnwidth,angle=0,keepaspectratio=true]
                  {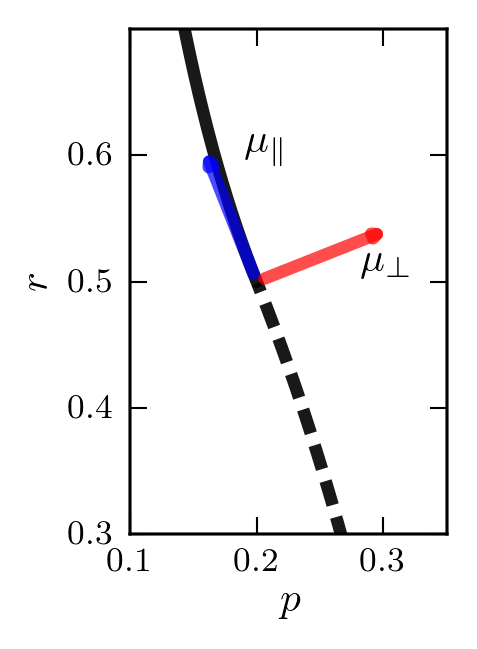}
		}
	\end{overpic}
	\caption{Phase diagram of the $\k=(2,3)$ mixture, showing the
          total mass of the percolating {\hkcore} cluster at different
          compositions $r$, for ER networks with $z_{1}=10$. The TCP
          at $r=1/2$ separates a line of \nth{1} order transitions
          (dashed) from the \nth{2} order line (solid). The masses of the
          $2$-rich-core (blue) and the $3$-rich-core (red) in the
          giant {\hkcore} are also shown. The inset shows the phase
          diagram in the $(r,p)$ space.}
	\label{fig:ph_diag_2_3}
\end{figure}%

For $k_a=1, k_b\geqslant 3$ mixtures on the Bethe lattice, the phase
diagram shows a critical line which meets a first order line at a
critical end point and a critical point at the end of a two-phase
coexistence between a low and a high density phase
\cite{baxter2011}.
The two lines do not match at a TCP, because the 1-nodes are so robust that a
1-rich phase is stable to damage at intermediate compositions even when the
$k_b$-rich phase has collapsed.
%
%
%
%
Let us consider now the case $\k=(2,3)$, with a
degree distribution such that $\sum_qq^2P(q)<\infty$.  We can rewrite
$Z$ (Eq. \ref{eq:Z-equation}) as $pf(Z) = 1\label{eq:practical-Z-eq}$
where
\begin{eqnarray}
\lefteqn{f(Z) = r\case{2P(2)}{\langle q\rangle} + \sum_{q\ge
3}\case{qP(q)}{\langle q\rangle} \times}\nonumber\\
& \times & \left[ \case{1-(1-Z)^{q-1}}{Z} - (1-r)(q-1)(1-Z)^{q-2}\right]
\end{eqnarray}
and similarly rewriting Eq.~\ref{eq:X-equation} as $h(X,Z) = 1/p$ with
\begin{eqnarray}
\lefteqn{h(X,Z) = r\case{2P(2)}{\langle q\rangle} + \sum_{q\ge
3}\case{qP(q)}{\langle q\rangle} \times}\nonumber\\
& \times & \left[ \case{1-(1-X)^{q-1}}{X} - (1-r)(q-1)(1-Z)^{q-2}\right].
\end{eqnarray}
These two equations differ only in the first (fractionary) part of the
sum. The $X$-dependent (positive) general term of
the series is monotonically decreasing for any $0<X\leqslant
1$, meaning that Eq.~(\ref{eq:X-equation}) has only one non-zero solution
when Eq.~(\ref{eq:Z-equation}) has a non-zero solution and therefore
$X=Z$ for the $\k=(2,3)$ mixture (and $S_{23}=M_{23}$). We expect this
property to be true for any mixture with nodes of type $k\geqslant2$.

We now explicitly show that the $\k=(2,3)$ mixture presents a TCP for
an {\er} (ER) degree distribution with mean degree $z_1$ $P(q)=z_{1}^{q}
\exp(-z_{1})/q!$. Using the condition $X=Z$, the equation
$pf(Z) = 1$ fully solves the problem of finding the onset of the giant
{\hkcore}, and the function $f(Z)$ becomes
$f(Z) =  \{1 - e^{-z_{1}Z} \left[ 1+ (1-r) z_{1} Z\right] \}/Z$.
It is now clear that $f'(Z)<0$ for
every $r>\frac{1}{2}$, implying that the only solution is the trivial
one $Z=0$, with a de-percolating \nth{2} order phase transition
occurring at the critical occupancy probability $p_c= 1/rz_1$. For
$r<\frac{1}{2}$, $f(Z)$ has a maximum at $0<Z_M<1$. This implies the
presence of a \nth{1} order transition and a coexistence between a
{\hkcore} phase of strength $M_{23}(Z_M)$, given by (\ref{eq:mass-k}),
and the non percolating phase at $Z=0$. The expansion of $f(Z)$ for $r
\geqslant \frac{1}{2}$, $Z(p)\to 0$, as $p\to p_c^+$, yields
$f(Z) = rz_1 + (1/2 - r)z_1^2Z + O\left(Z^2\right)$
showing that the maximum of $f(Z)$ continuously matches the $Z=0$
line exactly at $r_t=\frac{1}{2}$, where a TCP is present. We show the
computed phase diagram of the $\k=(2,3)$ mixture in
Fig.~\ref{fig:ph_diag_2_3}.

\begin{figure}[bht!]
\begin{minipage}{0.49\columnwidth}
\begin{overpic}[width=\columnwidth,grid=False]
{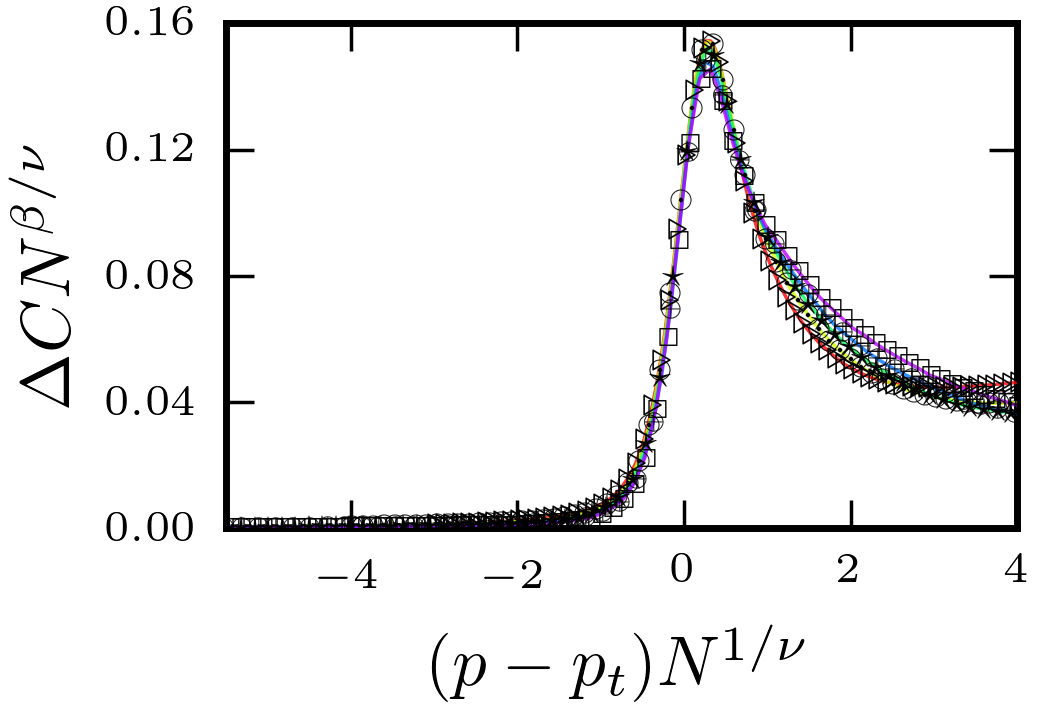}
\put(24,39){\includegraphics[width=0.35\columnwidth]
{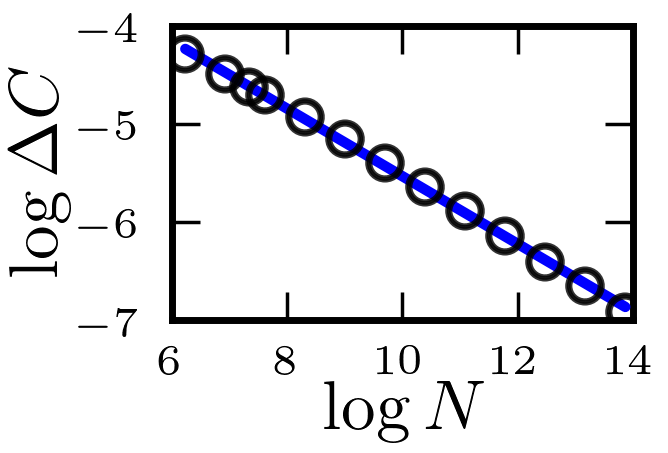}
}
\end{overpic}%
\end{minipage}
\begin{minipage}{0.49\columnwidth}
\begin{overpic}[width=\columnwidth,grid=False]
{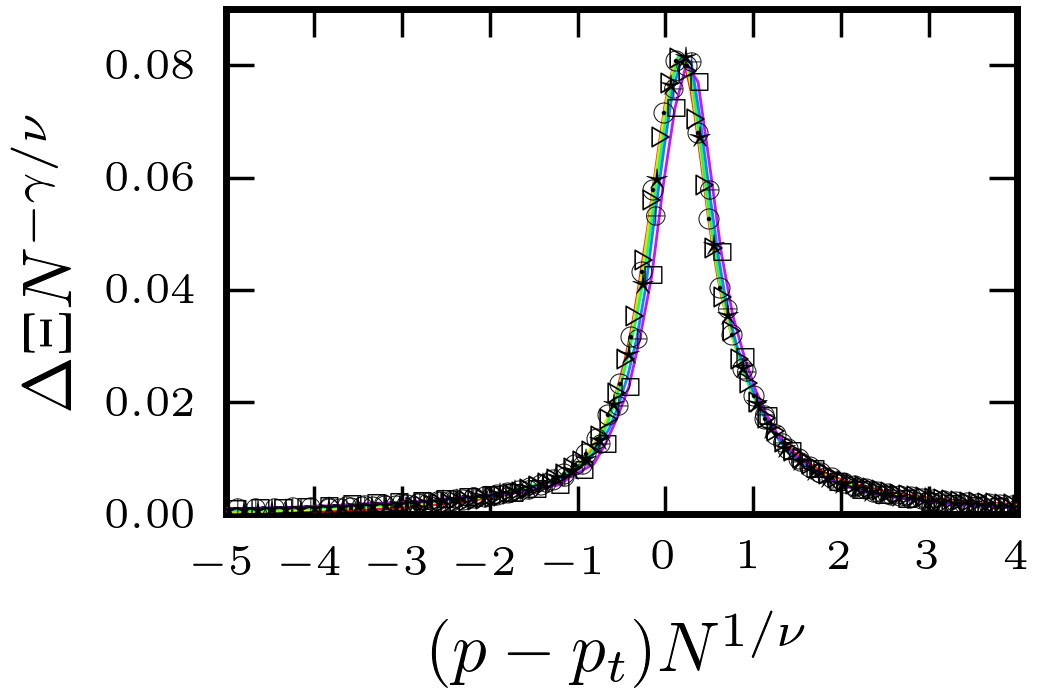}
\put(24,37){\includegraphics[width=0.35\columnwidth]
{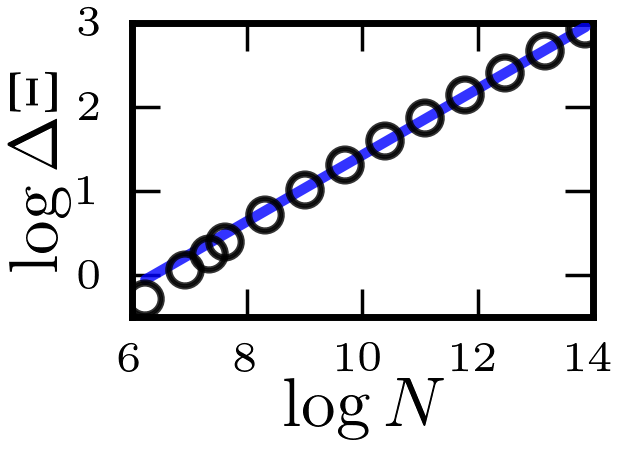}
}
\end{overpic}
\end{minipage}
\caption{Re-scaling of the corona mass $C_{23}$ and the mean corona cluster size
$\Xi_{23}$
  at the TCP on ER networks. The data range in size
  from $N=2^9$ to $N=2^{18}$ via successive doublings. We find the
  exponent ratios $\beta/\nu=0.34(5)$ and $\gamma/\nu=0.39(4)$ from the
  scaling of  $C_{23}$ and $\Xi_{23}$ at $p_{t}$ (insets) and show the data collapse achieved with those  exponents (main panels $N=2^{14}\dots 2^{18}$; $\triangleright, \odot, \star, \oplus, \Box$, respectively).}
\label{fig:scaling_deltas}
\end{figure}
We now calculate the critical exponents for this mixture, in
particular at the TCP at $r_t = \frac{1}{2}$, $p_t=\frac{2}{z_1}$. The expansion of
the order parameter $M_{23}(p)$ for $p\to p_c^+$ at $r \geqslant
\frac{1}{2}$ and $p\to p^{*+}$ (the border of the coexistence region) at $r < \frac{1}{2}$ yields three different values for the exponent $\beta$:
\begin{equation}
\beta = \left\{%
\begin{array}{lcrcl}
2 &&1/2 \leqslant &r& < 1\\
1 &&&r&=1/2\\
1/2  &&0 \leqslant &r&< 1/2
\end{array}
\right.
\end{equation}
The exponent $\beta$ takes a unique value at the TCP, and agrees with
the values found by Branco on the Bethe lattice
\cite{branco1993}.
However, in this work the presence of finite size cores had not been properly handled and it was erroneously assumed that the phase diagrams of the $\k=(1,3)$ and the $\k=(2,3)$ mixture  had the same topology.
The exponent $\beta=1/2$ for $r<1/2$ corresponds to
the usual hybrid phase transition seen in {\kcore} percolation, a discontinuous transition which combines with critical fluctuations (only on the percolating side) as usually found in \nth{2} order transitions.
To our knowledge, the $\k=(2,3)$ mixture is the first model displaying a TCP adjacent to a hybrid phase transition.

It has been shown that subsets of the HKC called corona clusters have the same critical properties of the HKC \cite{schwarz2006,goltsev2006}.
The corona vertices have exactly $k_{i}$ neighbours in the {\hkcore}, and form finite clusters whose mean size diverges when approaching the threshold from above.
The corona clusters provide a more convenient order parameter for
numerical study of the model on random networks, in contrast to the
{\hkcore} where only one (infinite) cluster survives. Using the
configuration model with ER degree distribution we simulated the
$\k=(2,3)$ mixture for various sizes.
The typical ansatz of finite size scaling for a continuous transition is that any quantity $Y$ scaling as $Y \sim (p-p_{c})^{-\chi}$ should
have the form
$Y = N^{\chi/\nu}F\left[(p - p_{c}) N^{1/\nu}\right]$,
where $\nu$ is the correlation length exponent and $F$ is some scaling
function. Given the universal nature of $F$ we expect to see data
collapse in a plot of $Y N^{-\chi/\nu}$ against $(p - p_{c})
N^{1/\nu}$. Computing the mass of the heterogeneous corona $C_{23}(k)$
at the TCP for various sizes we find $\beta/\nu =
0.34(5)$ (Fig.~\ref{fig:scaling_deltas}). Similarly for the mean corona
cluster size $\Xi_{23}$, we find $\gamma/\nu=0.39(4)$. We determine
the exponent $\nu=2.86(9)$ by the scaling of the effective percolation
threshold with size $p_{ave} - p_{c} \sim N^{-1/\nu}$, where we have
located $p_{ave}$ from the peak of the susceptibility of the corona
mass $\Delta C_{23} = (\left< C_{23} \right>^{2} - \left< C_{23}
\right>^{2})^{1/2}$. We find good data collapse with these exponents
in the scaling window at the TCP
(Fig.~\ref{fig:scaling_deltas}), and fit the exponents $\beta= 0.9(90)$
and $\gamma=1.13(1)$, the former being close to the value calculated
analytically.
The behavior of the strength of the {\hkcore} along the edges of the
coexistence region near the TCP for $r\to \frac{1}{2}^{+}$ allows
us to calculate analytically the subsidiary tricritical exponent $\beta_{u}$ defined
by $M^*(r) \sim \left(\frac{1}{2}-r\right)^{\beta_u}$
\cite{essam1971}. For $\k=(2,3)$ we find $\beta_{u}=2$.

The tricritical crossover exponent $\varphi_t$ describes the change of
the critical line as the TCP is approached \cite{riedel1972}.  Thus,
we write the critical line in terms of two scaling fields
$\mu_{\perp}$ and $\mu_{\parallel}$, perpendicular and tangent to the
critical line, respectively. Given the simplicity of the model, this
calculation can also be done analytically for ER networks. The rotation
defining the critical fields is
\begin{equation}
\left(
\begin{array}{c}
\mu_{\perp} \\
\mu_{\parallel}
\end{array} \right)
=
\left(
\begin{array}{cc}
\cos\vartheta & \sin\vartheta\\
-\sin\vartheta & \cos\vartheta \\
\end{array} \right)
\left(
\begin{array}{c}
p-p_t \\
r-r_t
\end{array} \right)
\end{equation}
with $\tan\vartheta = 4/z_1$.  Close to the TCP, the critical line has
a behavior $\mu_{\parallel}\sim \mu_{\perp}^{1/2}$, with a crossover
exponent $\varphi_t=2$ (Fig.~\ref{fig:ph_diag_2_3}).  We expect that the above critical behavior
(as well as the values of the critical exponents) is reproduced by all
degree distributions with finite second moment.

\begin{figure}[th!]
\begin{tabular}{lr}
\begin{minipage}[]{0.85\columnwidth}
\begin{overpic}[width=\columnwidth,grid=False]
{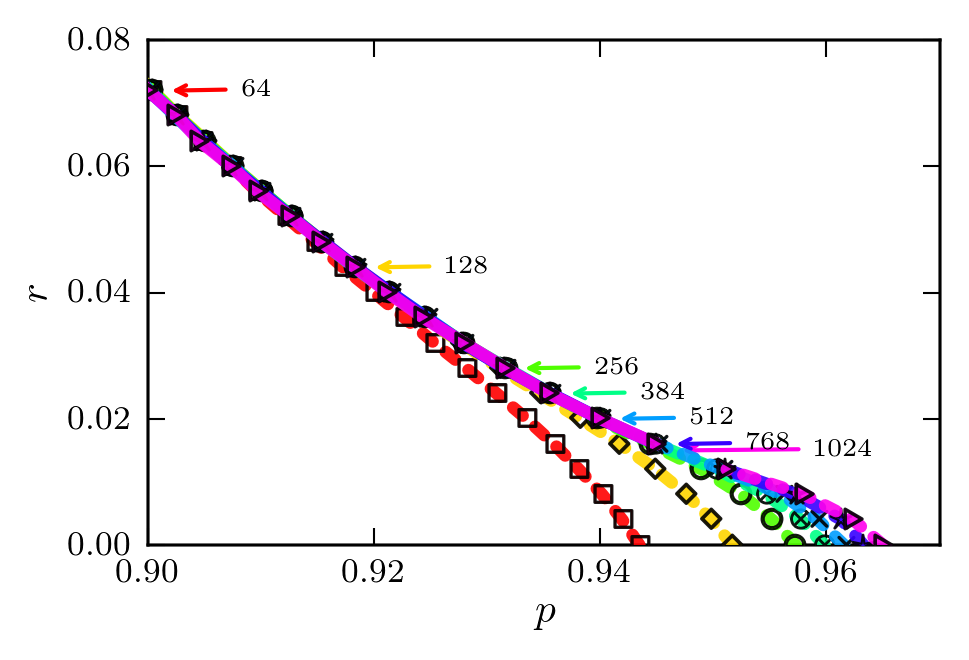}%
	\put(53,33){\includegraphics[width=0.43\columnwidth,keepaspectratio=true]%
          {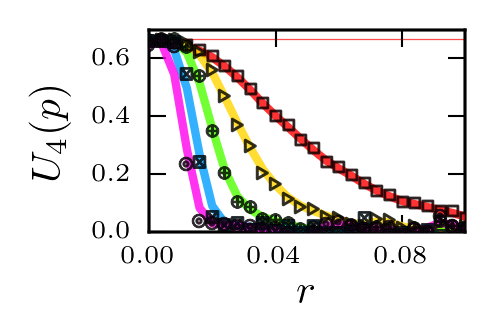}%
	}%
\end{overpic}
\end{minipage}%
\begin{minipage}[]{0.15\columnwidth}
\includegraphics[width=\columnwidth]%
{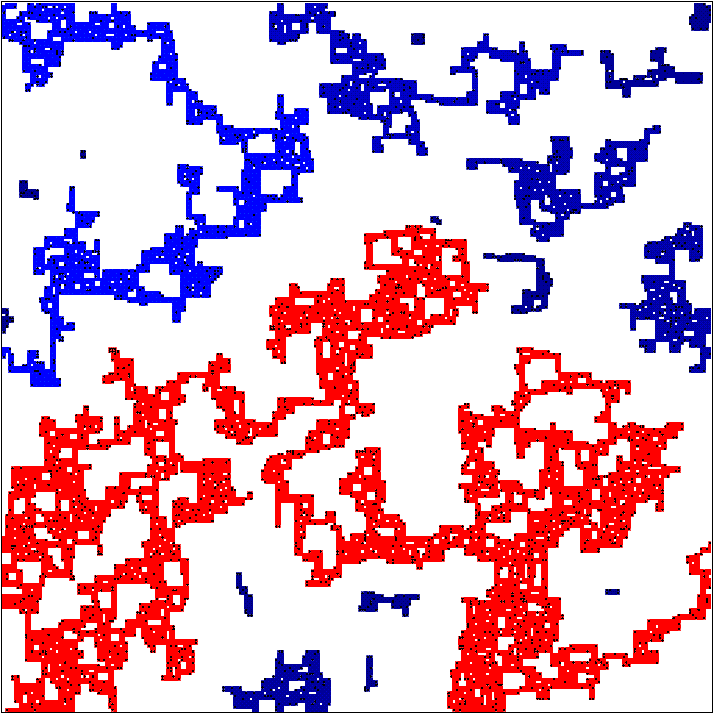}%
\label{fig:lattice_config_010}%
\\
\includegraphics[width=\columnwidth]%
{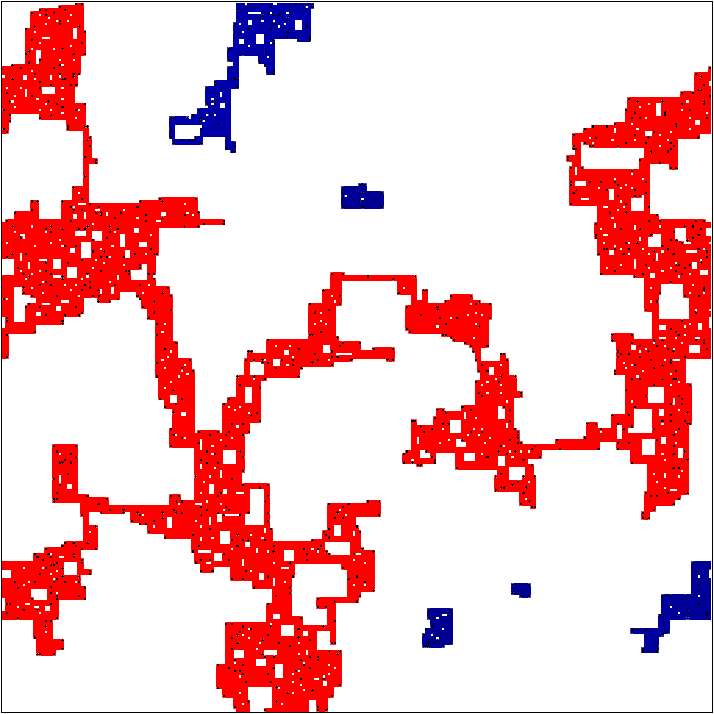}%
\label{fig:lattice_config_040}%
\\%
\includegraphics[width=\columnwidth]%
{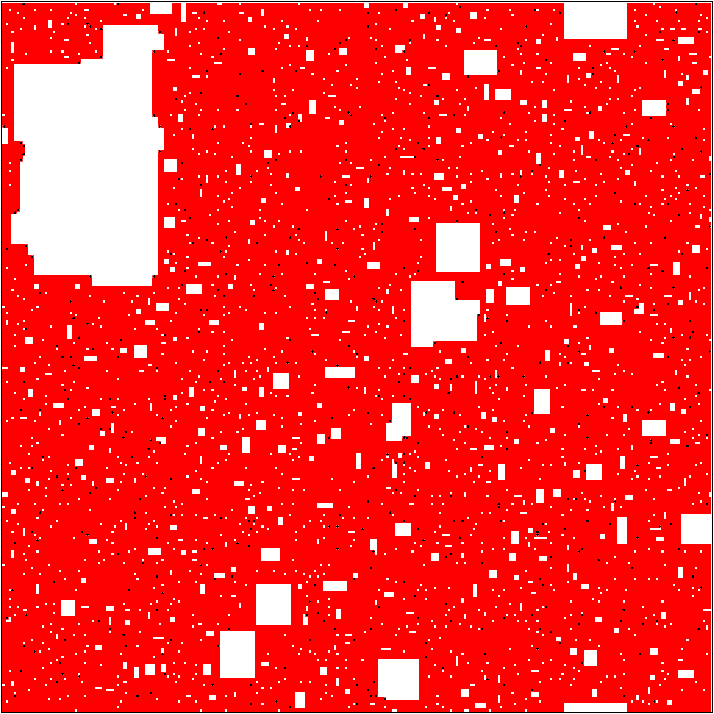}%
\label{fig:lattice_config_008}%
\vspace{15pt}
\end{minipage}
\end{tabular}
\caption{Phase diagram for the $\k=(2,3)$ lattice model showing the
  threshold density $p$ against composition $r$ for several sizes 
  $L= 64$($\Box$), 128($\diamond$), 256($\circ$), 384($\otimes$),
  512($\times$), 768($\star$), 1024($\triangleright$). The arrows
  indicate the location of the TCP for each $L$. The inset shows the
  Binder cumulant $U_{4}(p_{c})$ for various sizes, indicating the
  narrowing of the tricritical region for increasing size. On the
  right are sample configurations at the threshold density in the
  continuous transition region (top), at the TCP (centre) and in the
  discontinuous region (bottom); in each case the critical {\hkcore}
  cluster is red.}
\label{fig:phase_diagram_lattice}
\end{figure}

We simulated the $\k=(2,3)$ model on a 2d square lattice and located
the TCP at small concentrations of $k=2$ vertex types
(Fig.~\ref{fig:phase_diagram_lattice}). On the lattice, the analogous
bootstrap percolation (BP) model has been much studied and it is known
\cite{schonmann1992} that for $k\geqslant d+1$ a discontinuous
transition occurs only at $p=1$. For $k\leqslant d$ the transition is
continuous, although the critical exponents have values which in
general depend on $k$.
On the continuous side of the TCP we can expect the
usual scaling of the threshold density $p_{ave} -p_{c}\sim
L^{-1/\nu}$, whereas for $r=0$ we might expect the scaling form found
in BP $p_{ave}-p_{c} \sim 1/\log L$
\cite{aizenman1988}, although numerical
simulations have struggled to confirm this scaling in several cases
\cite{Adler:1989p5803,branco1999}. As shown in
Fig.~\ref{fig:phase_diagram_lattice} the TCP moves
toward $r=0$ with increasing size (determination of the precise
scaling with $L$ requires far larger sizes and is the subject of
further work). In fact, there is a finite window of $r$ over which the
transition slowly changes from \nth{1} to \nth{2} order, and this
window becomes sharper with increasing system size. We quantify this
with the Binder cumulant $U_{4}(p_{c}) = 1 - \left<M\right>^{4}/3
\left<M^{2}\right>^{2}$ which has the value $U_{4}=\frac{2}{3}$ on the
\nth{1} order side and $0$ on the \nth{2} order side (inset of Fig.~\ref{fig:phase_diagram_lattice}). Data collapse near the TCP does not work due to the presence of different scaling
regimes. We determined the critical exponents at the TCP
for the largest size simulated ($L=1024$) and found the exponents
$\beta=0.31(5)$, $\gamma=2.51(3)$ and $\nu=1.39(9)$.
Exponents $\gamma$ and $\nu$ at the TCP are very close to their values for ordinary
percolation on a 2d lattice (and a little smaller than the ones of explosive
percolation \cite{araujo2011}). Exponent $\beta$, instead, is
significantly larger. The fractal dimension of the tricritical
{\hkcore} clusters is $D=2-\beta/\nu=1.77(8)$, somewhat smaller than
ordinary percolation ($D=1.879$), reflecting the presence of
large, jagged voids in the $\k=(2,3)$ mixture at the TCP.
The unusual finite-size effects in this
model are reflected in a violation of the hyperscaling relation.

In contrast with the $\k=(2,3)$ case, the phase diagrams for $k_{a}=1$,
$k_{b}\geqslant3$ mixtures \cite{baxter2011} do not present a
TCP. Moreover, the analytical properties of $f(Z)$ and $h(X,Z)$ indicate that TCPs are also absent in mixtures of type
$k_a=2,k_b>3$.
Though far from ubiquitous, a TCP is indeed present in the $\k=(2,3)$
mixture, not only on the Bethe lattice but also in ER graphs and
regular square lattices.
This case appears to be peculiar because the resiliences of the 3-rich-phase and the 2-rich-phase are sufficiently close that the two phases collapse at the same damage fraction, leading to a complete failure of the \hkcore, either through a \nth{1} or a \nth{2} order transition.
This phenomenon may occur in cases when a mixing of a continuously and a discontinuously failing phase is not too heterogeneous ($k_b=k_a+1$).
It is intriguing to note that the case $\k=(2,3)$ almost exactly maps onto a
model of glasses, recently studied on the Bethe lattice \cite{sellitto2010},
which appears to be in the same universality class.


In conclusion, we have presented a new model of {\hkcore} percolation which supports
a smooth interpolation between classical percolation and a \nth{1}
order phase transition through a TCP. We are able to identify a new
tricritical scaling scenario and calculate, both by analytical and
numerical methods, critical exponents which are different from the
ones of known percolation transitions.  We prove the presence of this
critical phenomenon in ER graphs, and we also get strong numerical
evidence in the square lattice.
The capacity to govern the order of phase transitions in
randomly damaged networks may constitute a step towards a more
effective infrastructure for network protection.

We thank Hans Herrmann for useful comments. This work has been partially supported by Science Foundation Ireland under the grants 03/CE2/I303-1, 06/IN.1/I366 and 06/MI/005.

\begin{thebibliography}{10}%
\makeatletter
\providecommand \@ifxundefined [1]{%
 \ifx #1\undefined \expandafter \@firstoftwo
 \else \expandafter \@secondoftwo
\fi
}%
\providecommand \@ifnum [1]{%
 \ifnum #1\expandafter \@firstoftwo
 \else \expandafter \@secondoftwo
\fi
}%
\providecommand \enquote [1]{``#1''}%
\providecommand \bibnamefont  [1]{#1}%
\providecommand \bibfnamefont [1]{#1}%
\providecommand \citenamefont [1]{#1}%
\providecommand\href[0]{\@sanitize\@href}%
\providecommand\@href[1]{\endgroup\@@startlink{#1}\endgroup\@@href}%
\providecommand\@@href[1]{#1\@@endlink}%
\providecommand \@sanitize [0]{\begingroup\catcode`\&12\catcode`\#12\relax}%
\@ifxundefined \pdfoutput {\@firstoftwo}{%
 \@ifnum{\z@=\pdfoutput}{\@firstoftwo}{\@secondoftwo}%
}{%
 \providecommand\@@startlink[1]{\leavevmode}%
 \providecommand\@@endlink[0]{}%
}{%
 \providecommand\@@startlink[1]{%
  \leavevmode
  \pdfstartlink
   attr{/Border[0 0 1 ]/H/I/C[0 1 1]}%
   user{/Subtype/Link/A<</Type/Action/S/URI/URI(#1)>>}%
  \relax
 }%
 \providecommand\@@endlink[0]{\pdfendlink}%
}%
\providecommand \url  [0]{\begingroup\@sanitize \@url }%
\providecommand \@url [1]{\endgroup\@href {#1}{\urlprefix}}%
\providecommand \urlprefix [0]{URL }%
\providecommand \Eprint[0]{\href }%
\@ifxundefined \urlstyle {%
  \providecommand \doi [1]{doi:\discretionary{}{}{}#1}%
}{%
  \providecommand \doi [0]{doi:\discretionary{}{}{}\begingroup
  \urlstyle{rm}\Url }%
}%
\providecommand \doibase [0]{http://dx.doi.org/}%
\providecommand \Doi[1]{\href{\doibase#1}}%
\providecommand \bibAnnote [3]{%
  \BibitemShut{#1}%
  \begin{quotation}\noindent
    \textsc{Key:}\ #2\\\textsc{Annotation:}\ #3%
  \end{quotation}%
}%
\providecommand \bibAnnoteFile [2]{%
  \IfFileExists{#2}{\bibAnnote {#1} {#2} {\input{#2}}}{}%
}%
\providecommand \typeout [0]{\immediate \write \m@ne }%
\providecommand \selectlanguage [0]{\@gobble}%
\providecommand \bibinfo [0]{\@secondoftwo}%
\providecommand \bibfield [0]{\@secondoftwo}%
\providecommand \translation [1]{[#1]}%
\providecommand \BibitemOpen[0]{}%
\providecommand \bibitemStop [0]{}%
\providecommand \bibitemNoStop [0]{.\EOS\space}%
\providecommand \EOS [0]{\spacefactor3000\relax}%
\providecommand \BibitemShut [1]{\csname bibitem#1\endcsname}%
\bibitem{vespignani2010}%
  \BibitemOpen
  \bibfield{author}{%
  \bibinfo {author} {\bibfnamefont{A.}~\bibnamefont{Vespignani}},\ }%
  \bibfield{journal}{%
  \Doi{10.1038/464984a}{\bibinfo {journal} {Nature}}\ }%
  \textbf{\bibinfo {volume} {464}},\ \bibinfo {pages} {984} (\bibinfo {year}
  {2010})%
  \bibAnnoteFile{NoStop}{vespignani2010}%
\bibitem{araujo2011}%
  \BibitemOpen
  \bibfield{author}{%
  \bibinfo {author} {\bibfnamefont{N.~A.~M.}\ \bibnamefont{Ara\'{u}jo}},
  \bibinfo {author} {\bibfnamefont{J.~S.}\ \bibnamefont{Andrade}}, \bibinfo
  {author} {\bibfnamefont{R.~M.}\ \bibnamefont{Ziff}},\ and\ \bibinfo {author}
  {\bibfnamefont{H.~J.}\ \bibnamefont{Herrmann}},\ }%
  \bibfield{journal}{%
  \Doi{10.1103/PhysRevLett.106.095703}{\bibinfo {journal} {Phys. Rev. Lett.}}\
  }%
  \textbf{\bibinfo {volume} {106}},\ \bibinfo {pages} {095703} (\bibinfo {year}
  {2011})%
  \bibAnnoteFile{NoStop}{araujo2011}%
\bibitem{achlioptas2009}%
  \BibitemOpen
  \bibfield{author}{%
  \bibinfo {author} {\bibfnamefont{D.}~\bibnamefont{Achlioptas}}, \bibinfo
  {author} {\bibfnamefont{R.~M.}\ \bibnamefont{D'Souza}},\ and\ \bibinfo
  {author} {\bibfnamefont{J.}~\bibnamefont{Spencer}},\ }%
  \bibfield{journal}{%
  \Doi{10.1126/science.1167782}{\bibinfo {journal} {Science}}\ }%
  \textbf{\bibinfo {volume} {323}},\ \bibinfo {pages} {1453} (\bibinfo {year}
  {2009})%
  \bibAnnoteFile{NoStop}{achlioptas2009}%
\bibitem{dacosta2010}%
  \BibitemOpen
  \bibfield{author}{%
  \bibinfo {author} {\bibfnamefont{R.~A.}\ \bibnamefont{da~Costa}}, \bibinfo
  {author} {\bibfnamefont{S.~N.}\ \bibnamefont{Dorogovtsev}}, \bibinfo {author}
  {\bibfnamefont{A.~V.}\ \bibnamefont{Goltsev}},\ and\ \bibinfo {author}
  {\bibfnamefont{J.~F.~F.}\ \bibnamefont{Mendes}},\ }%
  \bibfield{journal}{%
  \Doi{10.1103/PhysRevLett.105.255701}{\bibinfo {journal} {Phys. Rev. Lett.}}\
  }%
  \textbf{\bibinfo {volume} {105}},\ \bibinfo {pages} {255701} (\bibinfo {year}
  {2010})%
  \bibAnnoteFile{NoStop}{dacosta2010}%
\bibitem{grassberger2011}%
  \BibitemOpen
  \bibfield{author}{%
  \bibinfo {author} {\bibfnamefont{P.}~\bibnamefont{Grassberger}}, \bibinfo
  {author} {\bibfnamefont{C.}~\bibnamefont{Christensen}}, \bibinfo {author}
  {\bibfnamefont{G.}~\bibnamefont{Bizhani}}, \bibinfo {author}
  {\bibfnamefont{S.~W.}\ \bibnamefont{Son}},\ and\ \bibinfo {author}
  {\bibfnamefont{M.}~\bibnamefont{Paczuski}},\ }%
  \bibfield{journal}{%
  \Doi{10.1103/PhysRevLett.106.225701}{\bibinfo {journal} {Phys. Rev. Lett.}}\
  }%
  \textbf{\bibinfo {volume} {106}},\ \bibinfo {pages} {225701} (\bibinfo {year}
  {2011})%
  \bibAnnoteFile{NoStop}{grassberger2011}%
\bibitem{riordan2011}%
  \BibitemOpen
  \bibfield{author}{%
  \bibinfo {author} {\bibfnamefont{O.}~\bibnamefont{Riordan}}\ and\ \bibinfo
  {author} {\bibfnamefont{L.}~\bibnamefont{Warnke}},\ }%
  \bibfield{journal}{%
  \Doi{10.1126/science.1206241}{\bibinfo {journal} {Science}}\ }%
  \textbf{\bibinfo {volume} {333}},\ \bibinfo {pages} {322} (\bibinfo {year}
  {2011})%
  \bibAnnoteFile{NoStop}{riordan2011}%
\bibitem{radicchi2009}%
  \BibitemOpen
  \bibfield{author}{%
  \bibinfo {author} {\bibfnamefont{F.}~\bibnamefont{Radicchi}}\ and\ \bibinfo
  {author} {\bibfnamefont{S.}~\bibnamefont{Fortunato}},\ }%
  \bibfield{journal}{%
  \Doi{10.1103/PhysRevLett.103.168701}{\bibinfo {journal} {Phys. Rev. Lett.}}\
  }%
  \textbf{\bibinfo {volume} {103}},\ \bibinfo {pages} {168701} (\bibinfo {year}
  {2009})%
  \bibAnnoteFile{NoStop}{radicchi2009}%
\bibitem{parshani2010}%
  \BibitemOpen
  \bibfield{author}{%
  \bibinfo {author} {\bibfnamefont{R.}~\bibnamefont{Parshani}}, \bibinfo
  {author} {\bibfnamefont{S.~V.}\ \bibnamefont{Buldyrev}},\ and\ \bibinfo
  {author} {\bibfnamefont{S.}~\bibnamefont{Havlin}},\ }%
  \bibfield{journal}{%
  \Doi{10.1103/PhysRevLett.105.048701}{\bibinfo {journal} {Phys. Rev. Lett.}}\
  }%
  \textbf{\bibinfo {volume} {105}},\ \bibinfo {pages} {048701} (\bibinfo {year}
  {2010})%
  \bibAnnoteFile{NoStop}{parshani2010}%
\bibitem{dorogovtsev2006}%
  \BibitemOpen
  \bibfield{author}{%
  \bibinfo {author} {\bibfnamefont{S.~N.}\ \bibnamefont{Dorogovtsev}}, \bibinfo
  {author} {\bibfnamefont{A.~V.}\ \bibnamefont{Goltsev}},\ and\ \bibinfo
  {author} {\bibfnamefont{J.~F.~F.}\ \bibnamefont{Mendes}},\ }%
  \bibfield{journal}{%
  \Doi{10.1103/PhysRevLett.96.040601}{\bibinfo {journal} {Phys. Rev. Lett.}}\
  }%
  \textbf{\bibinfo {volume} {96}},\ \bibinfo {pages} {040601} (\bibinfo {year}
  {2006})%
  \bibAnnoteFile{NoStop}{dorogovtsev2006}%
\bibitem{baxter2011}%
  \BibitemOpen
  \bibfield{author}{%
  \bibinfo {author} {\bibfnamefont{G.~J.}\ \bibnamefont{Baxter}}, \bibinfo
  {author} {\bibfnamefont{S.~N.}\ \bibnamefont{Dorogovtsev}}, \bibinfo {author}
  {\bibfnamefont{A.~V.}\ \bibnamefont{Goltsev}},\ and\ \bibinfo {author}
  {\bibfnamefont{J.~F.~F.}\ \bibnamefont{Mendes}},\ }%
  \bibfield{journal}{%
  \Doi{10.1103/PhysRevE.83.051134}{\bibinfo {journal} {Phys. Rev. E}}\ }%
  \textbf{\bibinfo {volume} {83}},\ \bibinfo {pages} {051134} (\bibinfo {year}
  {2011})%
  \bibAnnoteFile{NoStop}{baxter2011}%
\bibitem{sellitto2010}%
  \BibitemOpen
  \bibfield{author}{%
  \bibinfo {author} {\bibfnamefont{M.}~\bibnamefont{Sellitto}}, \bibinfo
  {author} {\bibfnamefont{D.}~\bibnamefont{{De Martino}}}, \bibinfo {author}
  {\bibfnamefont{F.}~\bibnamefont{Caccioli}},\ and\ \bibinfo {author}
  {\bibfnamefont{J.~J.}\ \bibnamefont{Arenzon}},\ }%
  \bibfield{journal}{%
  \Doi{10.1103/PhysRevLett.105.265704}{\bibinfo {journal} {Phys. Rev. Lett.}}\
  }%
  \textbf{\bibinfo {volume} {105}},\ \bibinfo {pages} {265704} (\bibinfo {year}
  {2010})%
  \bibAnnoteFile{NoStop}{sellitto2010}%
\bibitem{chalupa1979}%
  \BibitemOpen
  \bibfield{author}{%
  \bibinfo {author} {\bibfnamefont{J.}~\bibnamefont{Chalupa}}, \bibinfo
  {author} {\bibfnamefont{P.~L.}\ \bibnamefont{Leath}},\ and\ \bibinfo {author}
  {\bibfnamefont{G.~R.}\ \bibnamefont{Reich}},\ }%
  \bibfield{journal}{%
  \Doi{10.1088/0022-3719/12/1/008}{\bibinfo {journal} {J. Phys. C}}\ }%
  \textbf{\bibinfo {volume} {12}},\ \bibinfo {pages} {L31} (\bibinfo {year}
  {1979})%
  \bibAnnoteFile{NoStop}{chalupa1979}%
\bibitem{Balogh:2006p4406}%
  \BibitemOpen
  \bibfield{author}{%
  \bibinfo {author} {\bibfnamefont{J.}~\bibnamefont{Balogh}}\ and\ \bibinfo
  {author} {\bibfnamefont{B.~G.}\ \bibnamefont{Pittel}},\ }%
  \bibfield{journal}{%
  \Doi{10.1002/rsa.20158}{\bibinfo {journal} {Random. Struct. Algor.}}\ }%
  \textbf{\bibinfo {volume} {30}},\ \bibinfo {pages} {257} (\bibinfo {year}
  {2006})%
  \bibAnnoteFile{NoStop}{Balogh:2006p4406}%
\bibitem{Adler:1989p5803}%
  \BibitemOpen
  \bibfield{author}{%
  \bibinfo {author} {\bibfnamefont{J.}~\bibnamefont{Adler}}, \bibinfo {author}
  {\bibfnamefont{D.}~\bibnamefont{Stauffer}},\ and\ \bibinfo {author}
  {\bibfnamefont{A.}~\bibnamefont{Aharony}},\ }%
  \bibfield{journal}{%
  \Doi{10.1088/0305-4470/22/7/008}{\bibinfo {journal} {J. Phys. A}}\ }%
  \textbf{\bibinfo {volume} {22}},\ \bibinfo {pages} {L297} (\bibinfo {year}
  {1989})%
  \bibAnnoteFile{NoStop}{Adler:1989p5803}%
\bibitem{branco1999}%
  \BibitemOpen
  \bibfield{author}{%
  \bibinfo {author} {\bibfnamefont{N.~S.}\ \bibnamefont{Branco}}\ and\ \bibinfo
  {author} {\bibfnamefont{C.~J.}\ \bibnamefont{Silva}},\ }%
  \bibfield{journal}{%
  \Doi{10.1142/S0129183199000711}{\bibinfo {journal} {Int. J. Mod. Phys. C}}\
  }%
  \textbf{\bibinfo {volume} {10}},\ \bibinfo {pages} {921} (\bibinfo {year}
  {1999})%
  \bibAnnoteFile{NoStop}{branco1999}%
\bibitem{wuchty2005}%
  \BibitemOpen
  \bibfield{author}{%
  \bibinfo {author} {\bibfnamefont{S.}~\bibnamefont{Wuchty}}\ and\ \bibinfo
  {author} {\bibfnamefont{E.}~\bibnamefont{Almaas}},\ }%
  \bibfield{journal}{%
  \Doi{10.1002/pmic.200400962}{\bibinfo {journal} {Proteomics}}\ }%
  \textbf{\bibinfo {volume} {5}},\ \bibinfo {pages} {444} (\bibinfo {year}
  {2005})%
  \bibAnnoteFile{NoStop}{wuchty2005}%
\bibitem{schwarz2006}%
  \BibitemOpen
  \bibfield{author}{%
  \bibinfo {author} {\bibfnamefont{J.~M.}\ \bibnamefont{Schwarz}}, \bibinfo
  {author} {\bibfnamefont{A.~J.}\ \bibnamefont{Liu}},\ and\ \bibinfo {author}
  {\bibfnamefont{L.~Q.}\ \bibnamefont{Chayes}},\ }%
  \bibfield{journal}{%
  \Doi{10.1209/epl/i2005-10421-7}{\bibinfo {journal} {Europhys. Lett.}}\ }%
  \textbf{\bibinfo {volume} {73}},\ \bibinfo {pages} {560} (\bibinfo {year}
  {2006})%
  \bibAnnoteFile{NoStop}{schwarz2006}%
\bibitem{Chatterjee2007}%
  \BibitemOpen
  \bibfield{author}{%
  \bibinfo {author} {\bibfnamefont{N.}~\bibnamefont{Chatterjee}}\ and\ \bibinfo
  {author} {\bibfnamefont{S.}~\bibnamefont{Sinha}},\ }%
  \bibfield{journal}{%
  \bibinfo {journal} {Prog. Brain Res.}\ }%
  \textbf{\bibinfo {volume} {168}},\ \bibinfo {pages} {145} (\bibinfo {year}
  {2007})%
  \bibAnnoteFile{NoStop}{Chatterjee2007}%
\bibitem{alvarez2007}%
  \BibitemOpen
  \bibfield{author}{%
  \bibinfo {author} {\bibfnamefont{J.~I.}\ \bibnamefont{Alvarez-Hamelin}}\ and\
  \bibinfo {author} {\bibfnamefont{A.}~\bibnamefont{Puglisi}},\ }%
  \bibfield{journal}{%
  \Doi{10.1103/PhysRevE.75.051302}{\bibinfo {journal} {Phys. Rev. E}}\ }%
  \textbf{\bibinfo {volume} {75}},\ \bibinfo {pages} {51302} (\bibinfo {year}
  {2007})%
  \bibAnnoteFile{NoStop}{alvarez2007}%
\bibitem{klimek2009}%
  \BibitemOpen
  \bibfield{author}{%
  \bibinfo {author} {\bibfnamefont{P.}~\bibnamefont{Klimek}}, \bibinfo {author}
  {\bibfnamefont{S.}~\bibnamefont{Thurner}},\ and\ \bibinfo {author}
  {\bibfnamefont{R.}~\bibnamefont{Hanel}},\ }%
  \bibfield{journal}{%
  \Doi{10.1016/j.jtbi.2008.09.030}{\bibinfo {journal} {J. Theor. Biol.}}\ }%
  \textbf{\bibinfo {volume} {256}},\ \bibinfo {pages} {142} (\bibinfo {year}
  {2009})%
  \bibAnnoteFile{NoStop}{klimek2009}%
\bibitem{cohen2000}%
  \BibitemOpen
  \bibfield{author}{%
  \bibinfo {author} {\bibfnamefont{R.}~\bibnamefont{Cohen}}, \bibinfo {author}
  {\bibfnamefont{K.}~\bibnamefont{Erez}}, \bibinfo {author}
  {\bibfnamefont{D.}~\bibnamefont{ben Avraham}},\ and\ \bibinfo {author}
  {\bibfnamefont{S.}~\bibnamefont{Havlin}},\ }%
  \bibfield{journal}{%
  \Doi{10.1103/PhysRevLett.85.4626}{\bibinfo {journal} {Phys. Rev. Lett.}}\ }%
  \textbf{\bibinfo {volume} {85}},\ \bibinfo {pages} {4626} (\bibinfo {year}
  {2000})%
  \bibAnnoteFile{NoStop}{cohen2000}%
\bibitem{gleeson2008}%
  \BibitemOpen
  \bibfield{author}{%
  \bibinfo {author} {\bibfnamefont{J.~P.}\ \bibnamefont{Gleeson}},\ }%
  \bibfield{journal}{%
  \Doi{10.1103/PhysRevE.77.046117}{\bibinfo {journal} {Phys. Rev. E}}\ }%
  \textbf{\bibinfo {volume} {77}},\ \bibinfo {pages} {46117} (\bibinfo {year}
  {2008})%
  \bibAnnoteFile{NoStop}{gleeson2008}%
\bibitem{kitsak2010}%
  \BibitemOpen
  \bibfield{author}{%
  \bibinfo {author} {\bibfnamefont{M.}~\bibnamefont{Kitsak}}, \bibinfo {author}
  {\bibfnamefont{L.~K.}\ \bibnamefont{Gallos}}, \bibinfo {author}
  {\bibfnamefont{S.}~\bibnamefont{Havlin}}, \bibinfo {author}
  {\bibfnamefont{F.}~\bibnamefont{Liljeros}}, \bibinfo {author}
  {\bibfnamefont{L.}~\bibnamefont{Muchnik}}, \bibinfo {author}
  {\bibfnamefont{H.~E.}\ \bibnamefont{Stanley}},\ and\ \bibinfo {author}
  {\bibfnamefont{H.~A.}\ \bibnamefont{Makse}},\ }%
  \bibfield{journal}{%
  \Doi{10.1038/nphys1746}{\bibinfo {journal} {Nat. Phys.}}\ }%
  \textbf{\bibinfo {volume} {6}},\ \bibinfo {pages} {888} (\bibinfo {year}
  {2010})%
  \bibAnnoteFile{NoStop}{kitsak2010}%
\bibitem{goltsev2006}%
  \BibitemOpen
  \bibfield{author}{%
  \bibinfo {author} {\bibfnamefont{A.~V.}\ \bibnamefont{Goltsev}}, \bibinfo
  {author} {\bibfnamefont{S.~N.}\ \bibnamefont{Dorogovtsev}},\ and\ \bibinfo
  {author} {\bibfnamefont{J.~F.~F.}\ \bibnamefont{Mendes}},\ }%
  \bibfield{journal}{%
  \Doi{10.1103/PhysRevE.73.056101}{\bibinfo {journal} {Phys. Rev. E}}\ }%
  \textbf{\bibinfo {volume} {73}},\ \bibinfo {pages} {56101} (\bibinfo {year}
  {2006})%
  \bibAnnoteFile{NoStop}{goltsev2006}%
\bibitem{newman2003}%
  \BibitemOpen
  \bibfield{author}{%
  \bibinfo {author} {\bibfnamefont{M.~E.~J.}\ \bibnamefont{Newman}},\ }%
  \bibfield{journal}{%
  \bibinfo {journal} {SIAM Rev.}\ }%
  \textbf{\bibinfo {volume} {45}},\ \bibinfo {pages} {167} (\bibinfo {year}
  {2003})%
  \bibAnnoteFile{NoStop}{newman2003}%
\bibitem{branco1993}%
  \BibitemOpen
  \bibfield{author}{%
  \bibinfo {author} {\bibfnamefont{N.~S.}\ \bibnamefont{Branco}},\ }%
  \bibfield{journal}{%
  \Doi{10.1007/BF01053606}{\bibinfo {journal} {J. Stat. Phys.}}\ }%
  \textbf{\bibinfo {volume} {70}},\ \bibinfo {pages} {1035} (\bibinfo {year}
  {1993})%
  \bibAnnoteFile{NoStop}{branco1993}%
\bibitem{DeGregorio:2004p1889}%
  \BibitemOpen
  \bibfield{author}{%
  \bibinfo {author} {\bibfnamefont{P.}~\bibnamefont{{De Gregorio}}}, \bibinfo
  {author} {\bibfnamefont{A.}~\bibnamefont{Lawlor}}, \bibinfo {author}
  {\bibfnamefont{P.}~\bibnamefont{Bradley}},\ and\ \bibinfo {author}
  {\bibfnamefont{K.~A.}\ \bibnamefont{Dawson}},\ }%
  \bibfield{journal}{%
  \Doi{10.1103/PhysRevLett.93.025501}{\bibinfo {journal} {Phys. Rev. Lett.}}\
  }%
  \textbf{\bibinfo {volume} {93}},\ \bibinfo {pages} {25501} (\bibinfo {year}
  {2004})%
  \bibAnnoteFile{NoStop}{DeGregorio:2004p1889}%
\bibitem{essam1971}%
  \BibitemOpen
  \bibfield{author}{%
  \bibinfo {author} {\bibfnamefont{J.~W.}\ \bibnamefont{Essam}}\ and\ \bibinfo
  {author} {\bibfnamefont{K.~M.}\ \bibnamefont{Gwilym}},\ }%
  \bibfield{journal}{%
  \Doi{10.1088/0022-3719/4/10/015}{\bibinfo {journal} {J. Phys. C}}\ }%
  \textbf{\bibinfo {volume} {4}},\ \bibinfo {pages} {L228} (\bibinfo {year}
  {1971})%
  \bibAnnoteFile{NoStop}{essam1971}%
\bibitem{riedel1972}%
  \BibitemOpen
  \bibfield{author}{%
  \bibinfo {author} {\bibfnamefont{E.~K.}\ \bibnamefont{Riedel}},\ }%
  \bibfield{journal}{%
  \Doi{10.1103/PhysRevLett.28.675}{\bibinfo {journal} {Phys. Rev. Lett.}}\ }%
  \textbf{\bibinfo {volume} {28}},\ \bibinfo {pages} {675} (\bibinfo {year}
  {1972})%
  \bibAnnoteFile{NoStop}{riedel1972}%
\bibitem{schonmann1992}%
  \BibitemOpen
  \bibfield{author}{%
  \bibinfo {author} {\bibfnamefont{R.~H.}\ \bibnamefont{Schonmann}},\ }%
  \bibfield{journal}{%
  \Doi{10.1214/aop/1176989923}{\bibinfo {journal} {Ann. Prob.}}\ }%
  \textbf{\bibinfo {volume} {20}},\ \bibinfo {pages} {174} (\bibinfo {year}
  {1992})%
  \bibAnnoteFile{NoStop}{schonmann1992}%
\bibitem{aizenman1988}%
  \BibitemOpen
  \bibfield{author}{%
  \bibinfo {author} {\bibfnamefont{M.}~\bibnamefont{Aizenman}}\ and\ \bibinfo
  {author} {\bibfnamefont{J.~L.}\ \bibnamefont{Lebowitz}},\ }%
  \bibfield{journal}{%
  \Doi{10.1088/0305-4470/21/19/017}{\bibinfo {journal} {J. Phys. A}}\ }%
  \textbf{\bibinfo {volume} {21}},\ \bibinfo {pages} {3801} (\bibinfo {year}
  {1988})%
  \bibAnnoteFile{NoStop}{aizenman1988}%
\end{thebibliography}
%

\end{document}